\documentclass[letterpaper,titlepage,12pt]{article}

\pdfoutput=1

\usepackage{amssymb,amsmath,amsfonts}
\usepackage{graphicx}
\usepackage{epsfig}
\usepackage{ccaption}

\def\clock{{\count0=\time
           \divide\count0 60
           \ifnum\count0<10 0\fi\the\count0
           \multiply\count0 -60 \advance\count0 \time
           :\ifnum\count0<10 0\fi \the\count0
         }}
\newcommand{\timestamp}{{\small\vbox{\hbox{\tt\jobname.tex}
\hbox{\the\day/\the\month/\the\year, \clock}}}}

\newcommand{\nn}{\nonumber}
\newcommand{\ie}{{\it i.e.,\,}}
\newcommand{\eg}{{\it e.g.,\,}}

\newcommand{\lp}{\left(}
\newcommand{\rp}{\right)}
\newcommand{\vep}{\varepsilon}
\newcommand{\mc}[1]{\mathcal{#1}}

\newcommand{\beq}{\begin{equation}}
\newcommand{\eeq}{\end{equation}}
\newcommand{\bea}{\begin{eqnarray}}
\newcommand{\eea}{\end{eqnarray}}
\newcommand{\beqa}{\begin{eqnarray}}
\newcommand{\eeqa}{\end{eqnarray}}

\numberwithin{equation}{section}

\begin{document}

\begin{titlepage}
\begin{flushright}
\end{flushright}

\vskip 2cm
\begin{center}
{\bf\LARGE{Black Branes in a Box:
}}
\vskip 0.2cm
{\bf\LARGE{Hydrodynamics, Stability, and Criticality
}}
\vskip 1.2cm
{\bf 
Roberto Emparan$^{a,b}$,
Marina Mart{\'\i}nez$^{b}$
}
\vskip 0.5cm
\medskip
\textit{$^{a}$Instituci\'o Catalana de Recerca i Estudis
Avan\c cats (ICREA)}\\
\textit{Passeig Llu\'{\i}s Companys 23, E-08010 Barcelona, Spain}\\
\smallskip
\textit{$^{b}$Departament de F{\'\i}sica Fonamental and}\\
\textit{Institut de
Ci\`encies del Cosmos, Universitat de
Barcelona, }\\
\textit{Mart\'{\i} i Franqu\`es 1, E-08028 Barcelona, Spain}\\

\vskip .2 in

\end{center}

\vskip 0.3in

\baselineskip 16pt
\date{}

\begin{center} {\bf Abstract} \end{center} 

We study the effective hydrodynamics of neutral black branes enclosed in a finite cylindrical cavity with Dirichlet boundary conditions. We focus on how the Gregory-Laflamme instability changes as we vary the cavity radius $R$. Fixing the metric at the cavity wall increases the rigidity of the black brane by hindering gradients of the redshift on the wall. In the effective fluid, this is reflected in the growth of the squared speed of sound. As a consequence, when the cavity is smaller than a critical radius the black brane becomes dynamically stable. The correlation with the change in thermodynamic stability is transparent in our approach. We compute the bulk and shear viscosities of the black brane and find that they do not run with $R$. We find mean-field theory critical exponents near the critical point.

\vskip 0.2cm

\noindent 
\end{titlepage} \vfill\eject

\setcounter{equation}{0}

\pagestyle{empty}
\small
\normalsize
\pagestyle{plain}
\setcounter{page}{1}

\newpage

\section{Introduction and conclusions}

The complex non-linear dynamics of black holes and black branes, governed by Einstein's equations, can in some regimes be efficiently captured by an effective theory for collective degrees of freedom. Some of these degrees of freedom are worldvolume embedding coordinates, associated to the elastic dynamics characteristic of brane-like objects, but there are also hydrodynamic degrees of freedom --- \eg\ pressure and velocity --- associated to horizon dynamics. The hydrodynamic approach dates back to \cite{membrane} and has been greatly extended following the fluid/gravity correspondence of \cite{Bhattacharyya:2008jc}. In the context of asymptotically flat black branes, it features within the effective blackfold theory of \cite{Emparan:2009cs,Emparan:2009at,Camps:2012hw}. A framework for relating all these approaches is presented in \cite{EHR}.

Here we are interested in using the hydrodynamic methods for investigating a peculiar effect of black branes: the classical instability of their horizons to the formation of ripples along their worldvolume, discovered by Gregory and Laflamme in \cite{Gregory:1993vy}. Refs.~\cite{Emparan:2009at,Camps:2010br} have shown how this phenomenon is very neatly captured in the hydrodynamic theory of blackfolds: it is simply an instability of fluctuations of the pressure of the effective black brane fluid, \ie\ a sound-mode instability. 
In this paper we investigate it further by introducing an additional parameter to gain control over the stability of the system. To do so, we place the black brane inside a finite cylindrical cavity of fixed radius $R$. Then we find the solution for a fluctuating black brane, with regularity conditions at the horizon, in the hydrodynamic limit. We analyze how the effective black brane theory, in particular its stability under hydrodynamic fluctuations, changes as the cavity radius $R$ is varied. This allows us to study several issues:

\paragraph{Correlated instabilities --- ghosts vs.\ tachyons.} One reason to expect that enclosing the black brane in a cavity should have an effect on its stability comes from the Correlated Stability Conjecture (CSC) \cite{Gubser:2000mm}, which links classical dynamical stability to local thermodynamical stability. Since it is known that the specific heat of the black brane in a cavity changes from negative for cavity radii greater than a critical value, $R>R_c$, to positive when $R<R_c$ \cite{York:1986it}, the validity of the conjecture requires that the Gregory-Laflamme instability of the black brane disappears at the critical radius $R=R_c$. 

To put our study in the right context, it is worth discussing the status and proper interpretation of the CSC. This has often been taken as the statement that
\begin{itemize} 
\item (CSC:) translationally invariant horizons have a tachyonic perturbation mode if and only if they are locally thermodynamically unstable.
\end{itemize}
Indeed, a large part of the studies of the CSC have focused on the presence or absence of a static, zero-mode perturbation of finite wavelength (a tachyon) that would mark the onset of the instability \cite{Reall:2001ag,Harmark:2007md}. In particular, ref.~\cite{Gregory:2001bd} studied the static zero-mode for the black brane in a cavity and showed that it disappears precisely when the cavity size reaches $R=R_c$. 

This form of the CSC, however, is incorrect: examples of black branes are known which are thermodynamically stable but nevertheless have tachyonic instabilities \cite{Friess:2005zp}. 
Instead, the local thermodynamical stability of black branes is more appropriately related to the presence of massless \textit{ghost} excitations, rather than tachyons. These two kinds of unstable modes are quite different. For excitations that in some range of wavenumbers $k$ have a dispersion relation approximately of the type
\beq\label{omek}
\omega^2=c^2 k^2 +m^2\,,
\eeq
we say we have a tachyon when $m^2<0$. The static zero-mode corresponds to $k=k_0=\sqrt{-m^2/c^2}$ and $\omega=0$. Instead, we say we have a ghost when $c^2<0$, and in particular a massless ghost has $\Omega=\textrm{Im}\,\omega=\sqrt{-c^2}k$.

The argument why thermodynamical instabilities of a translation-invariant horizon are connected to massless ghosts is simple  \cite{Buchel:2005nt,Friess:2005zp,Emparan:2009at}. A horizon that is translationally invariant can support perturbations of arbitrarily long wavelength. In the cases where the frequency of these perturbations vanishes as the wavelength diverges, they are hydrodynamic modes, which are either fluctuations of conserved quantities or Goldstone modes. Both of them feature in the local thermodynamics of the fluid. In the case of main interest to us here, the conserved quantity is the energy, and it is an old result (which we reproduce in sec.~\ref{sec:hydrostab} below) that small fluctuations in the energy density propagate along the fluid with squared velocity
\beq
v_s^2=\frac{s}{C_V}\,,
\eeq
where $s$ is the entropy density of the fluid and $C_V$ its specific heat at fixed volume. Obviously, a local thermodynamic instability, with $C_V<0$, results in unstable perturbations of wavenumber $k$ whose amplitude grows exponentially in time like $\exp (\sqrt{-v_s^2}\,kt)$. More generally, any local thermodynamic instability of the black brane gives rise to a long-wavelength, hydrodynamic instability. In the terms used above, we have a massless ghost with $c^2=v_s^2<0$.
 
Then, the CSC as stated above must be replaced by a statement of \textit{Correlated Hydrodynamic Stability}:

\begin{itemize} 
\item (CHS:) translationally invariant horizons have massless ghost excitations if and only if they are locally thermodynamically unstable. The ghost is a long-wavelength, low imaginary frequency, hydrodynamic instability of the horizon.
\end{itemize}

Since horizons are stable to fluctuations of very short wavelength\footnote{On very short scales the horizon is indistinguishable from Minkowski space, which in any healthy gravitational theory is stable at sufficiently short wavelengths.}, the ghost instability at small $k$ must disappear at some larger $k=k_{0}>0$, \ie\ $\textrm{Im}\,\omega(k_{0})=0$. If also $\textrm{Re}\,\omega(k_{0})=0$, then this is a zero-mode. In other words, a hydrodynamic ghost instability (and hence a local thermodynamic instability) of the horizon will typically be accompanied by a tachyonic zero-mode at finite $k$. This is indeed the case for the GL instability of neutral black branes (also when in a cavity). But the converse need not be true: a tachyonic instability need not turn into a hydrodynamic ghost instability at very long wavelengths\footnote{For instance, it can become a homogeneous tachyonic mode with $\textrm{Im}\,\omega(k=0)=\sqrt{-m^2}$.}, and hence need not be related to a local thermodynamic instability.

In this paper we construct explicitly the ghost, hydrodynamic unstable perturbation of the black brane in a cavity, and show that it turns into an oscillatory (damped) sound wave when $R<R_c$.

\paragraph{Increasing rigidity.} Our analysis of the effective hydrodynamic theory also gives a concrete intuitive picture of why and how the instability disappears as the cavity size is reduced. The squared speed of sound of the effective theory at finite $R$ is a monotonic increasing function of decreasing $R$. A larger speed of sound reflects a higher rigidity of the system. This comes about because fixing the metric on the cavity wall at finite $R$ makes it harder for the geometry to fluctuate, and in particular prevents the creation of worldvolume gradients of the redshift on the wall. In the hydrodynamic theory, these gradients have the effect of an acceleration of the fluid that opposes the creation of inhomogeneities along the worldvolume. This works to make the system more rigid and therefore less unstable, until the instability disappears.

\paragraph{Viscosities do not run with $R$.} The solution for the spacetime metric for a fluctuating black brane in a cavity of radius $R$, to first-derivative order in the fluctuations, allows us to compute the stress-energy tensor of the effective  fluid including dissipative effects. The values we obtain for the shear and bulk viscosities (and for their ratios to the entropy density) are \textit{the same at all values of $R$}. This is very likely related to a similar result obtained in the context of black branes in AdS$_5$ in \cite{Brattan:2011my}, and may be a feature of a larger class of black branes.

\paragraph{Spectrum and criticality.} The inclusion of dissipative terms in the fluid equations gives us an improved approximation for the spectrum of unstable modes. The dispersion curves show clearly that the instability weakens as the critical point is approached. Although the hydrodynamical theory cannot capture all the physics of the critical state, it nevertheless indicates that critical exponents are of mean-field theory type, a result which is borne out by the numerical computations of \cite{Gregory:2001bd}.

\bigskip

In the remainder of the paper we elaborate on all these points in detail. Sec.~\ref{sec:staticbrane} introduces the black brane in a cavity as a static system. The solution for its fluctuations to first-derivative order is discussed in sec.~\ref{sec:flucbrane}. This solution forms the basis for the study  in sec.~\ref{sec:hydrostab} of stability from a hydrodynamical perspective, and its connection to local thermodynamic stability. Sec.~\ref{sec:visco} computes the effective viscosities of the black brane in the cavity of radius $R$, and then uses them for obtaining the dispersion relation for unstable modes. We conclude with a brief discussion of the critical point and the appearance of mean-field critical exponents.

\section{Static black brane in a cylindrical cavity}
\label{sec:staticbrane}

\subsection{Geometry}

We write the metric of a black $p$-brane in $D=3+p+n$ spacetime dimensions in the form
\beq\label{bbrmetric}
ds^2=\left( -f(r) u_a u_b +P_{ab}\right) d\sigma^a d\sigma^b + \frac{dr^2} {f(r)}+r^2d\Omega_{(n+1)}
\eeq
with
\beq
f(r)=1-\frac{r_0^n}{r^n}\,,
\eeq
and where
\beq
P_{ab}=\eta_{ab}+u_a u_b
\eeq
is the projector onto spatial direcions orthogonal to the timelike vector $u^a$ with normalization $\eta_{ab}u^a u^b=-1$.

We put the black brane inside a cylindrical cavity bounded by a `wall' that extends along the brane worldvolume directions $\sigma^a$ and which, in the transverse directions, is a sphere $S^{n+1}$ at finite radius $r=R$. We denote quantities measured on the cavity wall with a caret. The metric induced on the wall is
\beq\label{indmetric}
\hat h_{\mu\nu}dx^\mu dx^\nu=\hat h_{ab}d\sigma^a d\sigma^b +R^2 d\Omega_{(n+1)}
\eeq
with
\beqa\label{indhab}
\hat h_{ab}&=& -f(R) u_a u_b +P_{ab}\nn\\
&=&-\hat u_a \hat u_b+\hat P_{ab}\,.
\eeqa
Indices of hatted tensors will be raised and lowered with this metric.
The velocity
\beq\label{hatu}
\hat u_a=\sqrt{f(R)}\,u_a
\eeq
is unit-normalized with respect to this metric by absorbing the redshift factor on the wall. The orthogonal projector, instead, is not modified,
\beq
\hat P_{ab}=P_{ab}\,,
\eeq
since in the geometry \eqref{bbrmetric} the spatial worldvolume directions do not suffer any gravitational deformation.

The geometry of the wall of the cavity is characterized by giving, in addition to the induced metric, the extrinsic curvature tensor
\beq\label{Thmunu}
\Theta_{\mu\nu}=-\frac{1}{2}\sqrt{f(R)}\,\partial_R \hat h_{\mu\nu}\,.
\eeq
Out of this we obtain the Brown-York quasilocal stress-energy tensor on the wall. Since we are only interested in the dynamics of the worldvolume, we only consider the components of the tensor along the directions $\sigma^a$, and we integrate them over the transverse $S^{n+1}$ of radius $R$. The result is that
\beqa\label{stTab}
\hat T_{ab}&=&\frac{\Omega_{n+1}}{8\pi
G}R^{n+1}\lp\Theta_{ab}-\hat h_{ab}\Theta\rp\nn
\\
&=&
\frac{\Omega_{n+1}}{8\pi G}\left[ -(n+1) R^n \sqrt{f(R)}\, \hat u_a \hat u_b
+\partial_R\lp R^{n+1}\sqrt{f(R)}\rp \hat P_{ab} \right]\,.
\eeqa

\subsection{Physical magnitudes}

Eq.~\eqref{stTab} is a perfect-fluid stress-energy tensor
\beq
\hat T_{ab}=\hat\vep\, \hat u_a \hat u_b +\hat P\, \hat P_{ab}
\eeq
with energy density and pressure
\beqa
\hat\vep&=& -\frac{\Omega_{n+1}}{8\pi G}(n+1) R^{n}\sqrt{f(R)}\,,\\
\hat P&=&-\hat\vep +\frac{\Omega_{n+1}}{8\pi G}R^{n+1}\partial_R\sqrt{f(R)}
=-\hat\vep +\frac{\Omega_{n+1}}{16\pi G}\frac{n r_0^n}{\sqrt{f(R)}}\,.\label{hatP}
\eeqa

In addition, we can assign an entropy density and temperature to the system
\beqa\label{entdens}
s&=&\frac{\Omega_{n+1}}{4\pi G}r_0^{n+1}\,,\\
\hat{\mc{T}}&=&\frac{n}{4\pi r_0\sqrt{f(R)}}\,.
\eeqa
The temperature is modified relative to its asymptotic value by the redshift factor at the wall, but the entropy density does not depend on $R$: it is obtained as $s=S/\hat V$, where the total entropy $S$ is computed from the horizon area, and the spatial volume $\hat V$ does not undergo any variation as $R$ changes.

The system satisfies the thermodynamic Euler relation
\beq\label{eulertherm}
\hat\vep+\hat P=\hat{\mc{T}} s\,,  
\eeq
and the first law
\beq\label{1stlaw}
d\hat\vep=\hat{\mc{T}} ds\,,
\eeq
for variations that keep fixed the cavity radius $R$. 

For the record, we note that when $R$ is allowed to vary, the first law becomes 
\beq
d\hat\vep=\hat{\mc{T}} ds-\sigma_w da_w
\eeq 
where the wall area-density $a_w$ and tension $\sigma_w$ are 
\beq\label{awsw}
a_w=\Omega_{n+1}R^{n+1}\,,\qquad 
\sigma_w=\frac{n}{16\pi G}\frac{1+f(R)}{R\sqrt{f(R)}}\\.
\eeq
When $R$ can vary the wall is regarded as a dynamical object, and one gets the coupled dynamics of the black brane/wall system. However, although this might be of interest, for the remainder of the paper we will regard the wall only as a non-dynamical boundary condition.

\subsection{No subtraction required}

When $R\to\infty$ both $\hat\vep$ and $\hat P$ diverge, owing to the non-compactness of the space. A simple remedy to this is to subtract the stress-energy tensor associated to a surface in Minkowski space with the same induced metric $\hat h_{\mu\nu}$. However, we do not need this for our purposes. The reason is not merely that we keep $R$ finite and thus divergences are absent. More important, the intrinsic worldvolume dynamics that we are interested in is not affected by the subtraction. A surface at constant $r=R$ in Minkowski spacetime has $\Theta_{ab}=0$ and the stress-energy tensor $T_{ab}^{(M)}$ comes entirely from the curvature $\Theta$ of the $S^{n+1}$ of radius $R$. Then
\beqa\label{TabM}
T_{ab}^{(M)}=\frac{\Omega_{n+1}}{8\pi G}(n+1)R^{n}\hat h_{ab}\,.
\eeqa
Since we keep $R$ fixed, this stress-energy tensor is of `vacuum-type', \ie proportional to the worldvolume metric $\hat h_{ab}$ and with constant energy density. This is inert: it lacks any hydrodynamic behavior, which is associated with a breakdown of local Lorentz invariance and the presence of inhomogeneities on the worldvolume. 

Therefore, the subtraction does not affect the hydrodynamics of the brane, and we shall not implement it.\footnote{It would affect, though, the system in which the wall is dynamical and $R$ varies along the worldvolume.}

\section{Fluctuating black brane}
\label{sec:flucbrane}

We promote the parameters $u_a$ and $r_0$ in the solution to worldvolume collective degrees of freedom, \ie\ slowly-varying functions of $\sigma^a$. The remaining parameter, $R$, is kept fixed. Following \cite{Bhattacharyya:2008jc}, to the now fluctuating metric \eqref{bbrmetric} we add correcting functions $f_{\mu\nu}$ such that the total metric
\beqa\label{flmet}
ds^2&=&\left( \eta_{ab}+\frac{r_0(\sigma)^n}{r^n} u_a(\sigma) u_b(\sigma) \right) d\sigma^a d\sigma^b + \frac{dr^2} {1-\frac{r_0(\sigma)^n}{r^n}}+r^2d\Omega_{(n+1)}
\nonumber\\
&&+f_{\mu\nu}dx^\mu dx^\nu
\eeqa
is a solution to the field equations. We fix the radial coordinate by choosing it to be orthogonal to the worldvolume and normalized to measure the $S^{n+1}$-area-radius. Then $f_{\Omega\mu}=0$.
Working to leading order in derivatives, the correcting functions
\beq\label{fmunu}
f_{\mu\nu}dx^\mu dx^\nu=f_{ab}d\sigma^a d\sigma^b +2f_{ar}d\sigma^a dr+f_{rr} dr^2\,,
\eeq
can be decomposed into $SO(p)$-algebraically-irreducible terms in the form
\begin{eqnarray}\label{svt}
f_{ab}&=&\theta u_a u_b\, {\sf s}_1(r)+\frac1{p}\theta P_{ab}\, {\sf s}_2(r)+a_{(a} u_{b)}\,{\sf v}_1(r)+\sigma_{ab}\,{\sf t}(r)\,,\nn\\
f_{ar}&=& \theta u_a\, {\sf s}_3(r)+a_{a}\, {\sf v}_2(r)\,,\\
f_{rr}&=& \theta \left(1-\frac{r_0^n}{r^n}\right)^{-1}\,{\sf s}_4(r)\,,\nn
\end{eqnarray}
where 
\beq
\theta =\nabla_a u^a\,,\qquad a_a=u^b\nabla_b u_a\,,\qquad 
\sigma_{ab}={P_a}^c{P_b}^d\nabla_{(c} u_{d)}-\frac{\theta}{p}P_{ab}
\eeq
are respectively the expansion, acceleration, and shear of the flow of $u^a$ in the metric $\eta_{ab}$. Since these terms are algebraically independent, each of the sets of functions, ${\sf s}_i$ (scalar sector), ${\sf v}_i$ (vector sector), and ${\sf t}$ (tensor sector) decouple from the others in the linearized equations and can be studied separately.

The Einstein equations ${R^r}_{a}=0$ do not involve the $f_{\mu\nu}$ and are independent of $r$. Thus they are `constraint equations', and can be written in the form
\beq\label{consteqs}
\nabla_{a}\ln r_0^{n+1}=\theta u_a +(n+1)a_a\,.
\eeq
These equations allow to eliminate the derivatives of $r_0(\sigma)$ in terms of velocity gradients. Below we will return to their interpretation in fluid-dynamical terms.

In order to specify boundary conditions at $r=R$, we demand that the induced metric remains fixed and uncorrected to the order we are working,
\beq
\hat h_{ab}=-\hat u_a \hat u_b +P_{ab}+O(\partial^2)\,.
\eeq
This requires that
\beq\label{bcon1}
{\sf s}_1(R)={\sf s}_2(R)={\sf v}_1(R)={\sf t}(R)=0\,.
\eeq
In addition, we ask that the stress-energy tensor is in `Landau frame', defined such that the corrections $\hat T^{(1)}_{ab}$ to the leading order value lie entirely along spatial directions, \ie
\beq
\hat u^a\hat T^{(1)}_{ab}=0\,.
\eeq
A brief calculation shows that this implies the conditions
\beq\label{bcon2}
{\sf v}_1'(R)=0\,,\qquad {\sf s}_2'(R)=\frac{n+1}{R}{\sf s}_4(R)\,.
\eeq

The construction of the solution to the Einstein equations for $f_{\mu\nu}$ that satisfies these boundary conditions and in addition is regular at the horizon, is done in appendix~\ref{app:soln} using the results of \cite{Camps:2010br}. The explicit results are in eqs.~\eqref{solnR}. This provides the complete metric for the fluctuating black brane, to first-derivative order, for any solution of the equations \eqref{consteqs}.

The solution, however, is written in terms of the velocity field $u_a$ and the connection $\nabla_a$ for the metric $\eta_{ab}$, which is not the physical metric on the wall at $r=R$.\footnote{It is neither the metric on the surface at $r\to\infty$, since with our boundary conditions the functions ${\sf s}_{1,2}$ and ${\sf t}$ do not vanish there.} Nevertheless, we can readily find the relation of the latter to quantities on the wall. We do this in appendix \ref{app:conn}, where we find that
\beqa\label{hatgrad}
\hat u^a&=&\frac{u^a}{\sqrt{f(R)}}\,,\qquad 
\hat\theta=\frac{\theta}{\sqrt{f(R)}}\,,\qquad
\hat\sigma_{ab}=\frac{\sigma_{ab}}{\sqrt{f(R)}}\,,\nonumber\\
\hat a_a&=&a_a+\frac{1}{\sqrt{f(R)}}{P_a}^b\partial_b\sqrt{f(R)}\,.
\eeqa
The change in the velocity, expansion and shear in \eqref{hatgrad} is simply a local redshift. The acceleration is not redshifted, but it is affected by the spatial variation of the redshift along the worldvolume. The point is clearer if we introduce the Newtonian potential $\phi$,
\beq
f(R)=e^{2\phi}\,,
\eeq
which depends on $\sigma^a$ through $r_0$. Its spatial gradient is
\begin{equation}\label{anabphi}
{\boldsymbol\nabla}_a \phi\equiv {P_a}^b\partial_b\phi\,,
\end{equation}
and we see that the modification of the acceleration is due to a `force' term,
\begin{equation}\label{ahata}
\hat a_a=  a_a+{\boldsymbol\nabla}_a \phi\,.
\end{equation}

Now using these relations we write the metric in terms of wall quantities as
\beqa\label{solwall}
ds^2&=&\biggl( 
-\frac{f(r)}{f(R)}\, \hat u_a \hat u_b 
	+ P_{ab} 
	+ \frac{\hat\theta}{\sqrt{f(R)}}\, \hat u_a \hat u_b\, {\sf s}_1(r)
	+ \frac{\sqrt{f(R)}}{p}\,\hat\theta P_{ab}\, {\sf s}_2(r) \nn\\
	&&\quad+ \frac{1}{\sqrt{f(R)}} ( \hat a-{\boldsymbol\nabla}\phi)_{(a} \hat u_{b)}\,{\sf v}_1(r)
	+ \sqrt{f(R)}\,\hat \sigma_{ab}\,{\sf t}(r)
\biggr) 
d\sigma^a d\sigma^b \nn\\
&&+ 2\lp \hat\theta\, \hat u_a\, {\sf s}_3(r)+ \lp \hat a_a-{\boldsymbol\nabla}_{a}\phi\rp\, {\sf v}_2(r)\rp d\sigma^a dr\nonumber\\
&&+ \frac{dr^2} {f(r)}\lp 1+ \sqrt{f(R)}\,\hat\theta\, {\sf s}_4(r)\rp
+r^2d\Omega_{(n+1)}\,.
\eeqa

As we will see in the next section, the most consequential effect is the modification of the acceleration.

\section{Hydrodynamics and stability}
\label{sec:hydrostab}

For later reference we review briefly some generic features of the dynamics of perfect fluids.

\subsection{Perfect fluid dynamics}

The hydrodynamic equations $\nabla_a T^{ab}=0$ for a generic relativistic perfect fluid with stress-energy tensor
\beq
T_{ab}=\vep\,u_a u_b +P\, P_{ab}
\eeq
are 
\beq\label{fleqs}
u^a u^b \nabla_b\vep +P^{ab}\nabla_b P +(\vep + P)(\theta u^a + a^a)=0\,.
\eeq
The fluid is assumed to satisfy the local thermodynamical relations
$\vep+P= Ts$ and $d\vep =Tds$. Defining also
\beq
v_s^2=\frac{dP}{d\vep}\,,
\eeq
we can write the fluid equations \eqref{fleqs} in a conveniently simple form
\beq\label{fleqs2}
\nabla_a\ln s=\theta u_a -\frac1{v_s^2}a_a\,.
\eeq
Consider now a fluid state initially in static homogeneous equilibrium in its rest frame, and introduce a small perturbation,
\beq
s\to s+\delta s\, e^{i\omega t+ i {\bf k}\cdot {\bf x}}\,,\qquad
u^a=(1,{\bf 0}) \to (1,\delta{\bf  u}\,e^{i\omega t+ i {\bf k}\cdot {\bf x}})\,.
\eeq
Then the solution to the linearized eqs.~\eqref{fleqs2} gives fluctuations with dispersion relation
\beq
\omega(k)= \sqrt{v_s^2}\; k+O( k^2)\,,
\eeq
where $k=|\mathbf{k}|$. Hence $v_s$ is the velocity of propagation of small density fluctuations, \ie\ the speed of sound.

\subsection{Black brane hydrodynamics}
\label{sec:bbhydro}

In the black brane fluid, the entropy density $s$ is directly related to the horizon thickness $r_0$ by \eqref{entdens}. Therefore, density fluctuations in the fluid are variations of the horizon radius.

We can immediately see the hydrodynamic Gregory-Laflamme instability in the simplest case in which the cavity wall is removed, $R\to\infty$ \cite{Emparan:2009at,Camps:2010br}. In this case the induced metric is $\eta_{ab}$, the effective fluid velocity is $u_a$ and, comparing to \eqref{fleqs2}, we see that the constraint eqs.~\eqref{consteqs} are the equations of the effective relativistic fluid at asymptotic infinity. The effective speed of sound is
\beq
v_s^2=-\frac{1}{n+1}\,,
\eeq
which is imaginary and therefore fluctuations of $r_0$  grow exponentially in time instead of oscillating as sound waves. This is the Gregory-Laflamme instability in the regime of long wavelengths and small (imaginary) frequencies.

\paragraph{Finite cavity: effect of redshift gradients.} When we insert the cavity wall at finite $R$, the gradient term in \eqref{ahata} modifies the acceleration with which the effective fluid responds to a change in $r_0$.
A local fluctuation $\delta r_0>0$ results in a smaller $\phi$, which tends to push the effective fluid \textit{away} from the region of increased $r_0$. Conversely, a region of locally smaller $r_0$ gives a gradient term that accelerates the fluid towards that region. 
Therefore, as a consequence of fixing the metric on the cavity wall, the creation of inhomogeneities along the worldvolume is hindered. The result is to make the fluid more stable. Moreover, the effect is more pronounced as the cavity radius $R$ gets closer to the brane, since the redshift becomes stronger.\footnote{Note that the effect is the opposite of what would occur to a material fluid localized on a brane at finite $R$: this would be gravitationally pulled \textit{towards} larger local mass densities, \ie\ larger $r_0$. Instead, our effective fluid is not any matter in the spacetime, but rather it is a `holographic' description of the black brane.} 

We can be more quantitative if we use \eqref{consteqs} to write
\beq
{P_a}^b\nabla_b\ln r_0=a_a
\eeq
and then
\beq\label{gradphi}
{\boldsymbol\nabla}_{a}\phi=-\frac{n}{2f(R)}\frac{r_0^{n-1}}{R^n}{P_a}^b\nabla_b r_0
=-\frac{n}{2}\lp\frac{1}{f(R)}-1\rp a_a\,.
\eeq
Since $f(R)<1$, we see that ${\boldsymbol\nabla}_{a}\phi$ is directed opposite to $a_a$ and therefore opposes the unstable growth of inhomogeneities. If, by decreasing $R$, the gradient grows to a value such that
\beq
{\boldsymbol\nabla}_{a}\phi=-a_a\,,
\eeq
then in this state the acceleration of the fluid on the wall vanishes, $\hat a_a=0$: the black brane does not react to a density fluctuation, and the instability disappears. This happens when
\beq
\frac{n}{2}\lp\frac{1}{f(R)}-1\rp=1\,,
\eeq
that is, when
\beq\label{Rcrit}
R=R_c=r_0\left(\frac{n+2}{2}\right)^{1/n}\,.
\eeq
If we reduce $R$ below $R_c$, the acceleration $\hat a_a$ will be directed against the inhomogeneities, and the black brane will be stable.

\paragraph{Effective fluid equations and speed of sound.} We can frame this discussion in more fluid-dynamical terms. From \eqref{gradphi}, the relation between the accelerations \eqref{ahata} in the black brane fluid is
\beq\label{as}
\hat a_a=a_a\frac{\hat v_s^2}{v_s^2}\,,
\eeq
where
\beqa\label{hvs2}
\hat v_s^2&=&-\frac{1}{n+1}\lp 1-\frac{n}{2}\lp\frac{1}{f(R)}-1 \rp\rp\nn\\
&=&-\frac{1}{n+1}\frac{1-(R_c/R)^n}{f(R)}
\,.
\eeqa

Now the constraint equations \eqref{consteqs} written in terms of the effective fluid velocity on the wall become
\beq\label{constwall}
\hat\nabla_{a}\ln r_0^{n+1}=\hat\theta \hat u_a -\frac{1}{\hat v_s^2}\hat a_a\,.
\eeq
Comparing to the general form of the perfect fluid equations \eqref{fleqs2}, we see that 	\eqref{constwall} are the equations $\hat\nabla_a \hat T^{ab}=0$
for the stress-energy tensor \eqref{stTab} and $\hat v_s^2=(d\hat P/d\hat\vep)_R$ is indeed the speed of sound. The quasilocal stress-energy tensor is known to be conserved on general grounds \cite{Brown:1992br}, so the result is not surprising. What we have done here is to see explicitly how these conservation equations emerge, on walls at finite $R$, from the Einstein constraint equations. 

The explicit form of eqs.~\eqref{constwall} is in any case very illustrative. They show clearly that, to the order we work, all the flow with $R$ of the black brane dynamics is due to the modified acceleration term. Since the entropy density is independent of $R$, the change of the effective fluid with $R$ can be fully accounted for by the change of  $\hat v_s^2$.

The sound velocity $\hat v_s$ is imaginary for large $R$, but it vanishes when the cavity reaches the critical radius $R_c$ in \eqref{Rcrit}, and then becomes real for cavity radii $R\in (r_0,R_c)$. This change in stability works in the direction expected from our argument above. 

Intuitively, the speed of sound is a measure of the rigidity of the system to worldvolume fluctuations. For very large cavities, the geometry is excessively soft, indeed `anti-rigid', to the point of being unstable to deformations. 
The cavity wall, by fixing the geometry at a finite distance from the black brane, increases its stiffness and can even render it stable when the cavity is small enough. In fact, $\hat v_s^2$ grows without bound as the wall approaches the horizon, thus making the effective fluid incompressible in that limit.\footnote{The fact that $\hat v_s\to\infty$ as $R\to r_0$ does not necessarily entail any violation of causality. Hydrodynamic fluctuations are low-frequency modes, and causality is controlled by modes in the high-frequency end of the spectrum, see \eg\ the discussion in \cite{Marolf:2012dr}.} 

The growing stiffness of the system caused by worldvolume gradients of the redshift is also apparent in the expressions for the extrinsic curvature, \eqref{Thmunu}, and the pressure, \eqref{hatP}. Thus, the effective hydrodynamic theory explains in a simple manner why and how the black brane turns from unstably soft to stably stiff.

\paragraph{Unstable perturbation.}

It is now easy to give the complete form of the unstable black brane solution in the cavity. We illustrate it, for simplicity, in the case of a black string with worldsheet coordinates $\sigma^a=(t,x)$. Take a velocity profile of the form
\beq
\hat u_t= -1\,,\qquad \hat u_x=\exp\lp\sqrt{-\hat v_s^2}\,k t\rp\cos(k x)\,\delta u \,,
\eeq
and work to linear order in the small amplitude $\delta u$. The metric is given by \eqref{solwall}, with the functions ${\sf s}_i(r)$, ${\sf v}_i(r)$ as in eqs.~\eqref{solnR}, and with
\beq
\hat P_{xx}= 1\,,\qquad \hat\sigma_{xx}=0\,,
\eeq
\beqa
\hat\theta&=&-k \exp\lp\sqrt{-\hat v_s^2}\,k t\rp \sin(k x)\,\delta u\,,\\
\hat a_x &=&\sqrt{-\hat v_s^2}\,k \exp\lp\sqrt{-\hat v_s^2}\,k t\rp\cos(k x)\,\delta u\,,
\eeqa
and 
\beq
\hat a_x -{\boldsymbol\nabla}_{x}\phi=\frac{v_s^2}{\hat v_s^2}\hat a_x=
\frac{f(R)}{1-(R_c/R)^n}\hat a_x
\,.
\eeq

\subsection{Correlated dynamical and thermodynamic stability}

We can easily see that the change in dynamical stability at $R=R_c$ corresponds precisely to the change in the local thermodynamic stability of the black brane, \ie\ in the thermodynamic stability of the black hole that one obtains at any given point on the worldvolume. Since \eqref{eulertherm} and \eqref{1stlaw} imply $d\hat P=sd\hat{\mc{T}}$ we have
\beq\label{CV}
\left(\frac{d\hat P}{d\hat\vep}\right)_R=\frac{s}{\hat C_V}
\eeq
where $\hat C_V$ is the specific heat at fixed volume. Since the sign of $\hat C_V$ determines the local thermodynamic stability, the connection between the latter and the dynamical stability of the brane, in the hydrodynamic regime, is obvious. This is nothing but the fact that stability of hydrodynamic modes associated to conserved quantities is governed by the local thermodynamic properties of the fluid. 

In the calculation of the speed of sound and in eq.~\eqref{CV} we only need the static brane solution. What our study of the fluctuating brane shows is that there is indeed an explicit solution for a black brane in a fixed cavity which is regular on the horizon and which is dynamically stable or unstable in accord with its thermodynamical stability.

\section{Viscous hydrodynamics}
\label{sec:visco}

Having the fluctuating black brane geometry to first order in velocity gradients, we extract its quasilocal stress-energy tensor at finite $R$ including dissipative terms.

\subsection{Bulk and shear viscosities do not run}

The general form of the stress-energy tensor on the cavity wall at finite $R$, in the spacetime given by \eqref{flmet}, \eqref{fmunu}, \eqref{svt}, and with boundary conditions \eqref{bcon1} and \eqref{bcon2}, is
\beqa\label{flTab}
\hat T_{ab}&=&
\frac{\Omega_{n+1}}{8\pi G}\left[  -(n+1) R^n \sqrt{f(R)}\, \hat u_a \hat u_b
+\partial_R\lp R^{n+1}\sqrt{f(R)}\rp \hat P_{ab} 
 \right]\nonumber\\
&&
-\hat\zeta\hat\theta \hat P_{ab}-2\hat\eta \hat\sigma_{ab}+O(\partial^2)\,,
\eeqa
with bulk and shear viscosities
\beq
\hat\zeta =\frac{\Omega_{n+1}}{8\pi G}\left[ \frac{R^{n+1}}{2}{\sf s}_1'(R)+ \left(\frac{n+1}{2p}(R^n-r_0^n)+ \frac{n}{4}r_0^n\right){\sf s}_4(R)\right]\,,
\eeq
and
\beq
\hat\eta = \frac{\Omega_{n+1}}{8\pi G}\frac{R^{n+1}}{4}f(R) {\sf t}'(R)\,.
\eeq
Substituting the explicit values for the solution that is regular on the horizon, we get
\beq
\hat\zeta =\frac{s}{2\pi}\lp\frac1p+\frac1{n+1}\rp 
\eeq
and
\beq
\hat\eta=\frac{s}{4\pi}\,,
\eeq
with $s$ the entropy density \eqref{entdens}.

While the result for the shear viscosity is not surprising, the fact that the bulk viscosity remains the same at all $R$ is probably less obviously expected. In particular, observe that it is of the form
\beq
\hat\zeta=\frac{s}{2\pi}\lp\frac1p-v_s^2\rp \,.
\eeq
Thus it depends on the asymptotic value of the speed of sound, instead of its value $\hat v_s^2$ at the cavity wall, which one might naively have guessed. Had it been the latter case, $\hat\zeta$ would have run with $R$. Since $s$ is independent of $R$, we can equivalently say that neither $\hat\eta/s$ nor $\hat\zeta/s$ run with $R$.

This absence of running of $\hat\zeta$ is most probably related to the one in \cite{Brattan:2011my}, where it was found that for AdS black branes in a finite cavity the bulk viscosity remains zero at all $R$, despite the fact that the wall breaks conformal invariance. There exists an explicit mapping between AdS gravity and the sector of vacuum gravity involved in our system \cite{camps} which is independent of the cavity wall. Conceivably, it relates our result to that of \cite{Brattan:2011my} and possibly makes clearer why $\hat\zeta$ depends on $v_s^2$ instead of $\hat v_s^2$.

For AdS black branes, ref.~\cite{Brattan:2011my} found an intriguing relation, $\hat T^a{}_a= -d\hat\vep/ d \ln R$, for the running of the energy density with $R$. For our neutral black branes, the same equation formally applies if we set $n=-p-1$, which is not any physical black brane, and in fact corresponds to setting $D=2$. Since, again formally, when $n=-p-1$ one gets $\hat\zeta=0$, this running is valid including first-derivative corrections. The reason why this result holds in this context is possibly related to properties of analytic continuation in $n$, but at present its ultimate meaning is unclear to us.

\subsection{Spectrum of unstable modes}

With $\hat\zeta$ and $\hat\eta$ we can compute the corrections to the fluid equations due to the viscous damping of density fluctuations. This gives us a better approximation for the spectrum of unstable modes at finite $R$.

\begin{figure}[t]
\centering%
\includegraphics[scale=1]{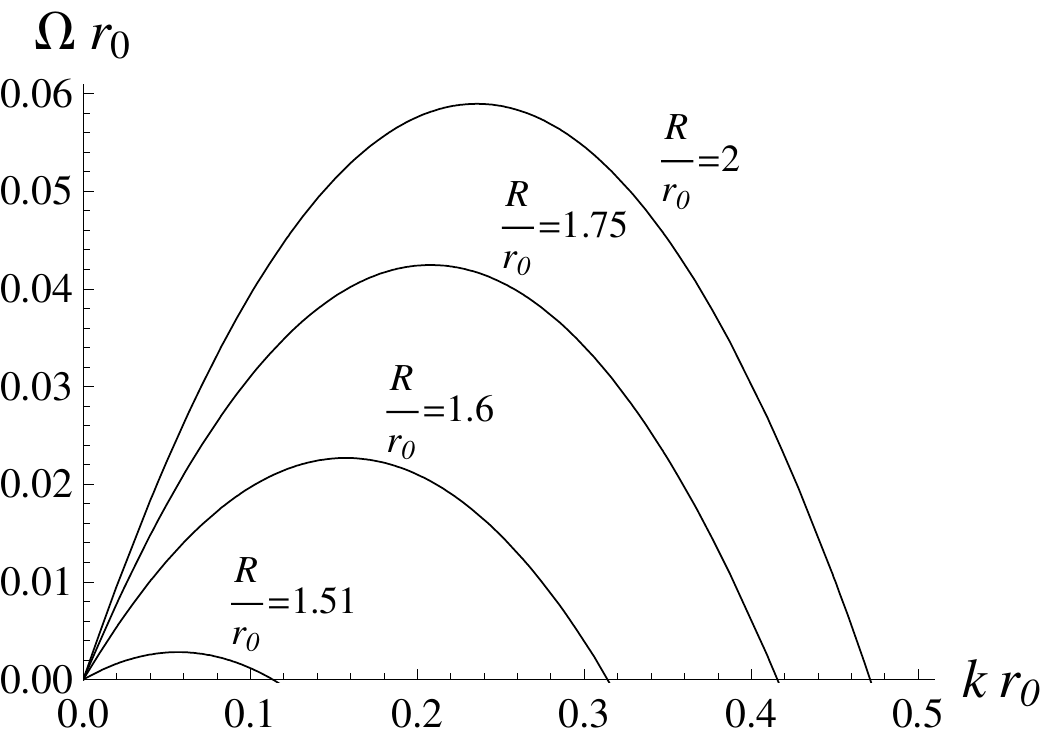}
\caption{\small Spectrum of GL unstable modes for a black brane in a cavity of radius $R$, according to eq.~\eqref{homk}. The curves correspond to $n=1$, for which the critical radius is $R_c=1.5\, r_0$.
}\label{fig:GLR}
\end{figure}

Solving the fluid equations to quadratic order in momenta $k$, the unstable modes of the black brane in a cavity with $R>R_c$ have imaginary frequency
\beqa
\Omega(k)&=&\sqrt{-\hat v_s^2}\;k-\frac{1}{2\hat{\mc{T}}s} \left[ \lp 1-\frac1p\rp 2\hat\eta+\hat\zeta\right] k^2+O(k^3)\nonumber\\
&=& \frac{k}{\sqrt{n+1}}\sqrt{\frac{1-(R_c/R)^n}{f(R)}} -k^2r_0 \frac{n+2}{n(n+1)}\sqrt{f(R)}+O(k^3),
\label{homk}\eeqa
where $R_c$ is given in \eqref{Rcrit}. We illustrate this result in fig.~\ref{fig:GLR}. As $R$ approaches $R_c$, the instability gets weaker, having both a smaller rate of growth $\Omega$ and a shorter range of unstable wavenumbers $k$.

While there is no previous calculation of black brane instabilities in a finite cavity that we can match these curves to, we can compare against the computation in ref.~\cite{Gregory:2001bd} of the wavenumber $k_{GL}$ of the zero-mode, for which $\Omega({k_{GL}})=0$. 
Our analytic expression \eqref{homk}, truncated to quadratic order, gives
\beqa\label{hkgl}
k_{GL}=\frac1{r_0}\frac{n\sqrt{n+1}}{n+2}\frac{\sqrt{1-(R_c/R)^n}}{f(R)}
\,.
\eeqa
\begin{figure}[t]
\begin{tabular}{cc}
\includegraphics[scale=.7]{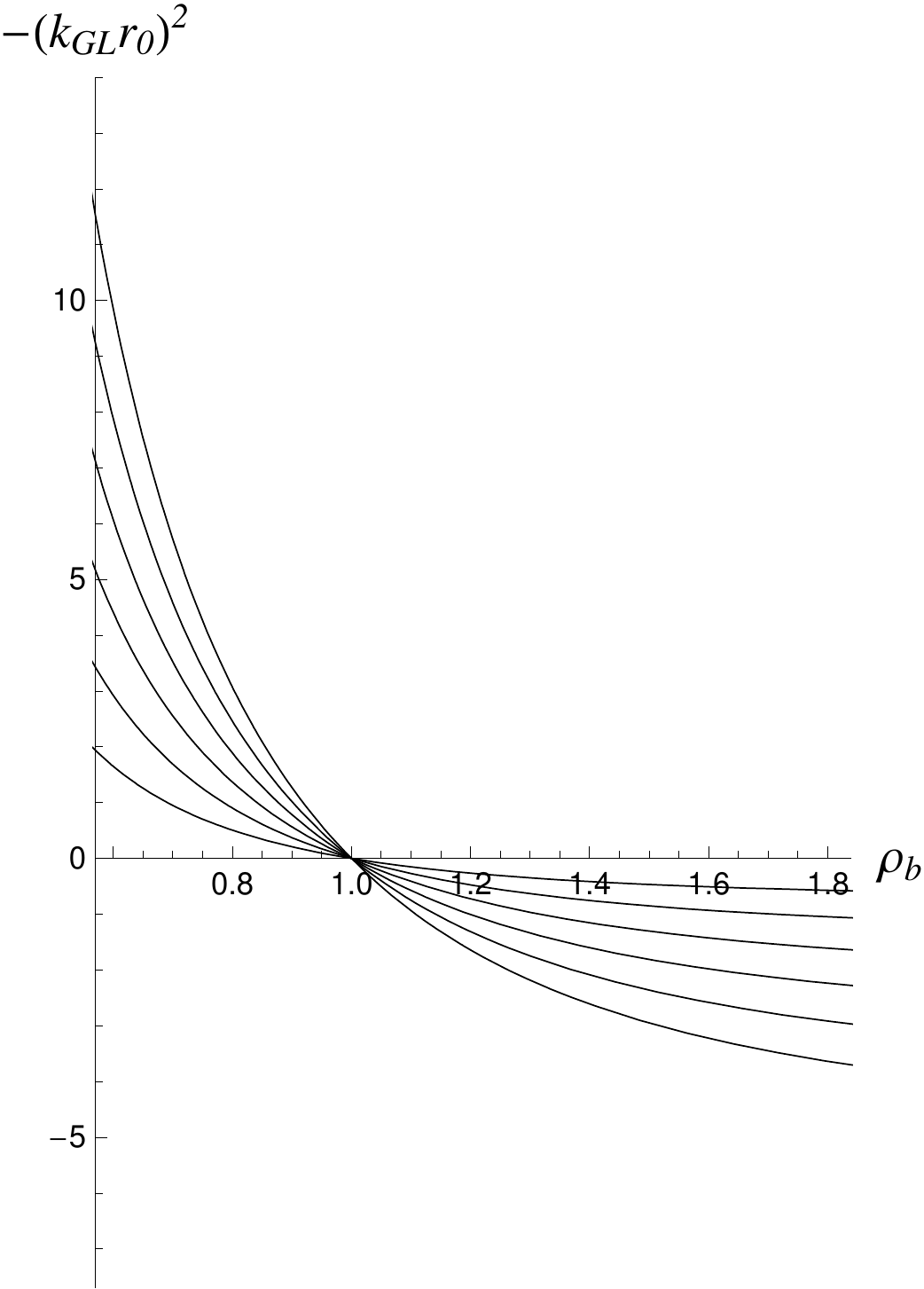}
\includegraphics[scale=.35]{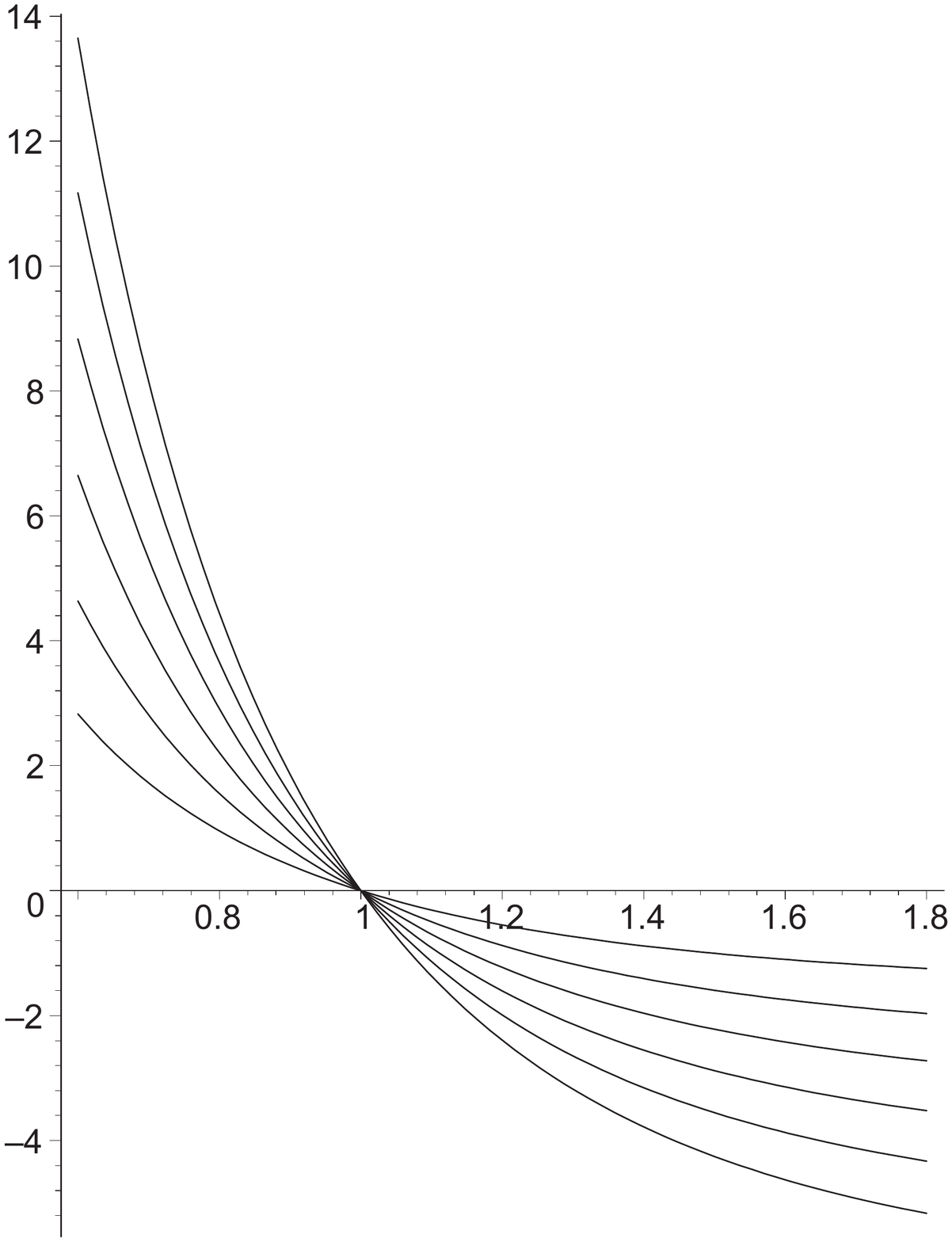}
\end{tabular}
\caption{\small Evolution of the GL zero mode $k_{GL}$ as a function of the cavity radius $R$. Left: obtained from the approximate analytical expression \eqref{hkgl}. Right: numerical results from \cite{Gregory:2001bd}. Following \cite{Gregory:2001bd}, in the vertical axis we display the `tachyon mass squared', $-(k_{GL}r_0)^2$, and in the horizontal axis $\rho_b=(R-r_0)/(R_c-r_0)$, so that the critical radius is always at $\rho_b=1$. The curves are for $n=2,\dots,7$, and lie closer to the horizontal axis the smaller $n$ is.  The finite, non-zero slope of the curves at $\rho_b=1$ indicates mean-field critical behavior \eqref{critk}.
}\label{fig:kGL}
\end{figure}
We display this result in fig.~\ref{fig:kGL}, where we compare it with the corresponding one of Fig.~3 in \cite{Gregory:2001bd}. The qualitative agreement between the two graphs is apparent, but one can easily discern quantitative discrepancies. These are expected, 
since \eqref{hkgl} has been obtained under the hydrodynamic assumption of small wavenumbers $k/\hat{\mc{T}} \ll 1$, which is not satisfied in general. As in \cite{Camps:2010br}, we may expect the agreement to improve for larger $n$.

\subsection{Critical behavior}

Eq.~\eqref{constwall} implies that when $\hat v_s^2=0$, the acceleration of the fluid under a density perturbation vanishes, \ie\ the fluid does not respond to variations of $r_0$. 
Using \eqref{CV} we see that at that point the effective specific heat $\hat C_V$ becomes infinite. This result makes manifest that the divergence of the specific heat is linked to the ghost instability --- the hydrodynamic mode is at the threshold of becoming ghost-like. The connection between the divergence of $\hat C_V$ and the tachyonic instability is, instead, only indirect: as discussed in the introduction, a hydrodynamic ghost instability of a black hole system is typically accompanied by a tachyonic instability. The hydrodynamic statement that $d\hat P/d\hat \vep\to 0$ means that the fluid does not react with any pressure gradient to local variations of the energy density. Thermodynamically, the fact that $\hat C_V\to\infty$ means that under these density variations the system does not create any temperature gradients that would restore it back to thermal equilibrium. Both effects, hydrodynamic and thermodynamic, are of course related via $d\hat P=sd\hat{\mc{T}}$. At the point where $\hat C_V\to\infty$, thermal fluctuations in the fluid have infinite wavelength. Then, this is a thermodynamic critical point.

In the critical state, the tachyonic zero mode becomes massless, \ie has infinite wavelength, and therefore it must be included alongside with the hydrodynamic modes in the effective low-frequency, long wavelength theory. Note, however, that this tachyon is never a proper hydrodynamic mode: the fluid equations never admit (to any arbitrary derivative order) a non-trivial static solution. The inclusion of the massless tachyon in the effective theory at the critical point cannot be done working solely within hydrodynamics.

Thus, the result \eqref{hkgl} from a truncated hydrodynamical calculation need not be accurate near the critical point. Note, however, that it predicts a critical behavior
\beq\label{critk}
k_{GL}\sim (R-R_c)^{1/2},
\eeq
which has a mean-field theory exponent. This appears to be correct: the numerical curves of \cite{Gregory:2001bd} for $-k_{GL}^2$, reproduced in the right-side graph in fig.~\ref{fig:kGL}, cut the horizontal axis with non-zero, finite slope consistently with the critical behavior \eqref{critk}. Perhaps we should not be too surprised: a mean-field theory description of the critical state might be natural in a classical field theory like general relativity and in a state with smooth horizons.

\paragraph{First order transition: inaccessibility of the critical point.}
The locally-unstable phases of the system of a black brane in a cavity are extremely fine-tuned states, which we have considered not so much for their possible relevance to actual physical phenomena but as being illustrative of the dynamics of black branes. But, actually, even the critical point and the associated second-order phase transition cannot be expected to be reached through any physical process, as they are dominated by a stronger first-order transition.

We have presented the problem starting from the black brane inside a very large cavity and then reduced the size of the cavity until the black brane becomes locally stable. However, from the perspective of a physical process it makes more sense to start from the system in the stable regime of $R\ll R_c$ and then follow it as we increase $R$. We are imagining that we keep the horizon size $r_0$ fixed, but one could easily consider other processes, \eg fix $R$ and change the temperature of the box. 

When $R$ reaches the value
\beq
R_{1}=\lp \frac{n+2}{2(n+1)}\rp^{1/n} R_c < R_c\,,
\eeq
the pressure $\hat P$ of the black brane becomes equal to that of Minkowski space in the same cavity, \eqref{TabM}, and for $R>R_1$ the latter has larger pressure. Thus, when the brane is `supercooled' at $R_{1}<R<R_c$, even if it is locally stable it will undergo a first-order phase transition and spontaneously nucleate bubbles of the `true vacuum', \ie hot flat space\footnote{For the black hole inside a spherical cavity ($p=0$) this is the analogue of the Hawking-Page transition.}. This will form holes in the worldvolume of the brane that then begin to expand. It may be interesting to study further this phenomenon, which shares features with the one studied in \cite{Horowitz:2007fe}. At any rate, it will prevent the observation of the critical state at $R=R_c$, and indeed of all the unstable black brane phases.

\section*{Acknowledgments}

We are greatly indebted to Joan Camps, Veronika Hubeny, and Mukund Rangamani, for sharing their insights on this problem in many discussions, and for their comments on the manuscript. We are also grateful to Simon Ross for providing and giving permission to use the right-side graph in fig.~\ref{fig:kGL}.
Work supported by MEC FPA2010-20807-C02-02, AGAUR 2009-SGR-168 and CPAN CSD2007-00042 Consolider-Ingenio 2010.

\appendix

\section{Solution with finite-cavity boundary conditions}
\label{app:soln}

Ref.~\cite{Camps:2010br} solved the Einstein equations for the perturbations to first order in derivatives and obtained the general solution that is regular on the future event horizon. The solution, expressed in Eddington-Finkelstein coordinates, contains a number of integration constants that are to be determined by conditions at the spatial boundary. For ease of comparison, we use the same notation for these constants ($c_{vr}$, $c_{ii}$, \textit{etc.}) as in \cite{Camps:2010br}. All the calculations at finite $R$, including the stress-energy tensor, could be carried out in Eddington-Finkelstein coordinates \cite{EHR}, but here we choose to work in Schwarzschild coordinates to maintain continuity with \cite{Camps:2012hw,Camps:2010br}.
In these coordinates the solution is
\begin{subequations}
\begin{align}
\mathsf{s}_1( r)&= \frac{n+2}{2(n+1)} \left(f(r)-\frac{n}{n+2}\right) \mathsf{s}_2(r)+ c_{vv}-2c_{vr} -f(r)c_{vv} \,,\displaybreak[0]\\
\mathsf{s}_2( r)&= \frac{2\,r_0}{n}\ln f(r)+c_{ii}\,,\displaybreak[0]\\
\mathsf{s}_3( r)&= \frac{1}{2f(r)}\left[\left(\frac{n}{n+1}-f(r)\right)
\mathsf{s}_2( r)+\frac{2r_0}{n+1}\lp f(r)-1\rp \left(n\frac{r_*}{r_0}+1\right)\right]\nonumber\\
&\quad+c_{vv}- c_{vr}-\frac{c_{vv}-2c_{vr}}{f(r)}
\,,\displaybreak[0]\\
\mathsf{s}_4( r)&= \lp f(r)^{-1}-1\rp\left( \frac{4 r_0-n \mathsf{s}_2(r)}{2(n+1)} + c_{vv}-2c_{vr}\right)\,,
\displaybreak[0]\\
\mathsf{v}_1( r)&=c_{vi}^{(2)}+\frac{c_{vi}^{(1)}}{r^n}\,,\displaybreak[0]\\
\mathsf{v}_2( r)&= \frac{r_*-r+\mathsf{v}_1( r)}{f(r)}+f_{rj}(r)\,,
\displaybreak[0]\\
\mathsf{t}(r)&= \mathsf{s}_2( r)+c_{ij}\,,
\end{align}
\end{subequations}
where
\beq
r_*=\int\frac{dr}{1-\frac{r_0^n}{r^n}}\,.
\eeq
The function $f_{rj}(r)$ is a gauge-dependent function only constrained to be finite on the horizon and thus could be set to zero. 

The boundary conditions \eqref{bcon1}, \eqref{bcon2} are satisfied by choosing
\beq
c_{vv}=c_{vr}=c_{vi}^{(1)}=c_{vi}^{(2)}=c_{ij}=0\,,\qquad
c_{ii}=-\frac{2\,r_0}{n}\ln f(R)\,,
\eeq 
with which
\begin{subequations}\label{solnR}
\begin{align}
\mathsf{s}_1( r)&= \frac{n+2}{2(n+1)} \left(f(r)-\frac{n}{n+2}\right) \mathsf{s}_2(r) \,,\displaybreak[0]\\
\mathsf{s}_2( r)&= \frac{2\,r_0}{n}\ln \frac{f(r)}{f(R)}\,,\displaybreak[0]\\
\mathsf{s}_3( r)&= \frac{1}{2f(r)}\left[\left(\frac{n}{n+1}-f(r)\right)
\mathsf{s}_2( r)+\frac{2r_0}{n+1}\lp f(r)-1\rp \left(n\frac{r_*}{r_0}+1\right)\right]
\,,\displaybreak[0]\\
\mathsf{s}_4( r)&= \lp f(r)^{-1}-1\rp\left( \frac{4 r_0-n \mathsf{s}_2(r)}{2(n+1)}\right)\,,
\displaybreak[0]\\
\mathsf{v}_1( r)&=0\,,\displaybreak[0]\\
\mathsf{v}_2( r)&= \frac{r_*-r}{f(r)}+f_{rj}(r)\,,
\displaybreak[0]\\
\mathsf{t}(r)&= \mathsf{s}_2( r)\,.
\end{align}
\end{subequations}
This is the unique solution, up to the gauge choice of $f_{rj}(r)$, that satisfies the regularity condition at the horizon and the boundary conditions on the wall at fixed $R$.

\section{Connection on the wall}
\label{app:conn}

We present here the relation between the connections $\nabla_a$ and $\hat\nabla_a$ compatible with, resp., the metrics $\eta_{ab}$ and $\hat h_{ab}$, and we use this to relate the gradients of the respective velocity vectors, $u_a$ and $\hat u_a$. We decompose these gradients, as usual, into traceless symmetric shear $\sigma_{ab}$, expansion $\theta$, acceleration $a_a$, and antisymmetric vorticity $\omega_{ab}$, so that
\beq
\nabla_a u_b=\sigma_{ab}+\frac1p \theta P_{ab}-u_a a_b+\omega_{ab}\,,
\eeq
and similarly for hatted quantities.

We follow the same steps as in \cite{Brattan:2011my}. If the difference between the metrics is 
\beq
\gamma_{ab}=\hat h_{ab}-\eta_{ab}
\eeq
then the difference between the connections, $\tilde\Gamma^c_{ab}$, such that
\beq
\hat\nabla_a V_b =\nabla_a V_b -\tilde\Gamma^c_{ab}V_c
\eeq
is given by
\beq
\tilde\Gamma^c_{ab}=\frac12 \hat h^{cd}\lp \nabla_a\gamma_{bd}+\nabla_b\gamma_{ad}-\nabla_d\gamma_{ab}\rp\,.
\eeq

In our case,
\beq
\gamma_{ab}=\lp 1-f\rp u_a u_b\,.
\eeq
Here we always take $f$ evaluated on the wall, \ie\ $f=f(R)$, which depends on $\sigma$ through $r_0(\sigma)$. 
We find
\beqa
\tilde\Gamma^c_{ab}&=&\frac{u^d\partial_d f}{2f}u^c u_a u_b-\frac{1}{f}u^cu_{(a}{P_{b)}}^d\partial_d f+ \frac12  u_a u_b P^{cd}\partial_d f\nonumber\\
&&+\lp f^{-1}-1\rp u^c\lp \sigma_{ab}+\frac1p \theta P_{ab}\rp + 2(1-f)u_{(a}{\omega_{b)}}^c-(1-f)a^c u_a u_b\,.
\eeqa

Using this and \eqref{hatu} we find that
\beq
\hat\nabla_a \hat u_b=\frac1{\sqrt{f}}\lp \sigma_{ab}+\frac1p \theta P_{ab}\rp + \sqrt{f}\omega_{ab}-\hat u_a \lp a_b+\frac1{2f} {P_b}^c\partial_c f \rp\,,
\eeq
from where we immediately deduce eqs.~\eqref{hatgrad} and 
\beq
\hat\omega_{ab}=\sqrt{f}\,\omega_{ab}\,.
\eeq 

Note that we have not made use anywhere of the fluid equations of motion for eliminating the derivatives of $f$ in favor of derivatives of the velocity. This is done in sec.~\ref{sec:bbhydro}

An alternative but equivalent way of obtaining the same results is the following. We may regard the two metrics as related, to zeroth derivative order, by the change $\hat u_a\hat\sigma^a =\sqrt{f(R)}\, u_a\sigma^a$. Above we have set $\hat u_a =\sqrt{f(R)}\, u_a$ and left the coordinates unchanged. But we could just as well leave $\hat u_a=u_a$, perform a coordinate rescaling (of time), and include the derivatives of $f(R)$ that result from this coordinate change into the correction terms $f_{\mu\nu}$.



\begin{thebibliography}{99}

\bibitem{membrane}
T. Damour, 
``Surface Effects in Black Hole Physics", Proceedings of the Second
Marcel Grossmann Meeting on General Relativity, (edited by R. Ruffini,
North Holland, 1982) p.\ 587.


\bibitem{Bhattacharyya:2008jc}
  S.~Bhattacharyya, V.~E.~Hubeny, S.~Minwalla and M.~Rangamani,
  ``Nonlinear Fluid Dynamics from Gravity,''
  JHEP {\bf 0802} (2008) 045
  [arXiv:0712.2456 [hep-th]].


\bibitem{Emparan:2009cs}
  R.~Emparan, T.~Harmark, V.~Niarchos and N.~A.~Obers,
  ``Worldvolume Effective Theory for Higher-Dimensional Black Holes (Blackfolds),''
  Phys.\ Rev.\ Lett.\  {\bf 102}, 191301 (2009)
  [arXiv:0902.0427 [hep-th]].


\bibitem{Emparan:2009at}
  R.~Emparan, T.~Harmark, V.~Niarchos and N.~A.~Obers,
  ``Essentials of Blackfold Dynamics,''
  JHEP {\bf 1003} (2010) 063
  [arXiv:0910.1601 [hep-th]].

\bibitem{Camps:2012hw}
  J.~Camps and R.~Emparan,
  ``Derivation of the blackfold effective theory,''
  JHEP {\bf 1203} (2012) 038
  [arXiv:1201.3506 [hep-th]].


\bibitem{EHR}
R.~Emparan, V.~E.~Hubeny and M.~Rangamani, to appear.

\bibitem{Gregory:1993vy}
  R.~Gregory and R.~Laflamme,
  ``Black strings and p-branes are unstable,''
  Phys.\ Rev.\ Lett.\  {\bf 70} (1993) 2837
  [hep-th/9301052].



\bibitem{Camps:2010br}
  J.~Camps, R.~Emparan and N.~Haddad,
  ``Black Brane Viscosity and the Gregory-Laflamme Instability,''
  JHEP {\bf 1005} (2010) 042
  [arXiv:1003.3636 [hep-th]].




\bibitem{Gubser:2000mm}
  S.~S.~Gubser and I.~Mitra,
  ``The Evolution of unstable black holes in anti-de Sitter space,''
  JHEP {\bf 0108} (2001) 018
  [hep-th/0011127].


\bibitem{York:1986it}
  J.~W.~York, Jr.,
  ``Black hole thermodynamics and the Euclidean Einstein action,''
  Phys.\ Rev.\ D {\bf 33} (1986) 2092.


\bibitem{Reall:2001ag}
  H.~S.~Reall,
  ``Classical and thermodynamic stability of black branes,''
  Phys.\ Rev.\ D {\bf 64} (2001) 044005
  [hep-th/0104071].

\bibitem{Harmark:2007md}
  T.~Harmark, V.~Niarchos and N.~A.~Obers,
  ``Instabilities of black strings and branes,''
  Class.\ Quant.\ Grav.\  {\bf 24} (2007) R1
  [hep-th/0701022].

\bibitem{Gregory:2001bd}
  J.~P.~Gregory and S.~F.~Ross,
  ``Stability and the negative mode for Schwarzschild in a finite cavity,''
  Phys.\ Rev.\ D {\bf 64} (2001) 124006
  [hep-th/0106220].

\bibitem{Buchel:2005nt}
  A.~Buchel,
  ``A Holographic perspective on Gubser-Mitra conjecture,''
  Nucl.\ Phys.\ B {\bf 731} (2005) 109
  [hep-th/0507275].

\bibitem{Friess:2005zp}
  J.~J.~Friess, S.~S.~Gubser and I.~Mitra,
  ``Counter-examples to the correlated stability conjecture,''
  Phys.\ Rev.\ D {\bf 72} (2005) 104019
  [hep-th/0508220].



\bibitem{Brattan:2011my}
  D.~Brattan, J.~Camps, R.~Loganayagam and M.~Rangamani,
  ``CFT dual of the AdS Dirichlet problem : Fluid/Gravity on cut-off surfaces,''
  JHEP {\bf 1112} (2011) 090
  [arXiv:1106.2577 [hep-th]].


\bibitem{Brown:1992br}
  J.~D.~Brown and J.~W.~York, Jr.,
  ``Quasilocal energy and conserved charges derived from the gravitational action,''
  Phys.\ Rev.\ D {\bf 47} (1993) 1407
  [gr-qc/9209012].

\bibitem{Marolf:2012dr}
  D.~Marolf and M.~Rangamani,
  ``Causality and the AdS Dirichlet problem,''
  JHEP {\bf 1204} (2012) 035
  [arXiv:1201.1233 [hep-th]].

\bibitem{camps}
M.~Caldarelli, J.~Camps, B.~Gouteraux, K.~Skenderis, private communication.
  
\bibitem{Horowitz:2007fe}
  G.~T.~Horowitz and M.~M.~Roberts,
  ``Dynamics of First Order Transitions with Gravity Duals,''
  JHEP {\bf 0702} (2007) 076
  [hep-th/0701099].


\end{thebibliography}
\end{document}